# Fabrication of Hydrophobic Inorganic Coatings on Natural Lotus Leaves for Nanoimprint Stamps

Shuxi Dai, Wanyong Ding, Yang Wang, Dianbo Zhang, Zuliang Du*

*Key Laboratory for Special Functional Materials of Ministry of Education, Henan University, Kaifeng 475004, PR China*

*Corresponding author. Tel:   +86-378-3881358.   FAX: +86-378-3881358.

 E-mail address:   zld@henu.edu.cn (Z. Du).

**Abstract**

 Hydrophobic inorganic films were obtained by direct deposition of copper or silicon onto natural lotus leaves by ion beam sputtering deposition technique. Scanning electron microscopy observations showed a lotus-leaf-like surface structure of the deposited inorganic films. Hydrophobic nature of the inorganic films on lotus leaves had been improved compared to the inorganic films deposited on flat silicon substrates. Water contact angles measured on the lotus-leaf-like copper and silicon films were $136.3 \pm 8°$ and $117.8 \pm 4.4°$, respectively. The hydrophobic lotus-leaf-like inorganic films had been repeated used as nanoimprint stamps. Negative structures of lotus-leaf-like inorganic films were obtained on the polystyrene resist layers.

## 1. Introduction

   Lotus leaves with super-hydrophobic and self-cleaning properties are of great interest for industrial applications and fundamental research. The hydrophobicity related to surface topography of lotus leaf consists of micrometer-scale papillae and branch-like wax nanostructures [1]. This unique hierarchical structure makes water droplets on the surface of lotus leaf rolling freely and removing dust and the hydrophobic mechanisms have been pointed out [2]. Many works focused on the fabrication of lotus-leaf-like surfaces to mimic the water-repellent properties of lotus leaves have been carried out. Super-hydrophobic lotus-leaf-like surface have been reported such as copper film [3-5], poly (vinyl chloride), polytetrafluoroethylene, and polystyrene surface [6-8].

Recently, soft lithography had been developed for the direct replication of lotus-leaf-like structures on polymeric and silica films using natural lotus leaf as template and polydimethylsiloxane (PDMS) as stamp [9-10]. Soft polydimethylsiloxane stamps were prepared by casting a liquid PDMS solution against the surface of the lotus leaf. After solidification for several hours at $80^o$, the PDMS layer was peeled off, resulting in a negative structure of the lotus leaf. Finally, positive lotus-leaf-like structures on polymeric and silica films were obtained by soft lithography with the PDMS stamps. However, a drawback of soft lithography method is that the low elastic modulus of soft PDMS stamps makes it difficult to repeated imprint due to sagging and structural collapse of the imprinted stamp [11]. Hard stamps with lotus-leaf-like structures are necessary to provide a repeated and large-scale preparation process of the unique micro/nano structures using high-throughput nanoimprint lithography.

In this paper, hard stamps of inorganic films were obtained by ion beam sputtering deposition of copper and silicon directly on natural lotus leaves templates. The coated inorganic films kept the papillary micro/nano structures and hydrophobic properties of lotus leaves. The inorganic metal and silicon deposition enhanced the strength of hard stamps and made it possible to be used as stamps in nanoimprint lithography. The repeated nanoimprint process had been performed using the hard stamps on the polystyrene resist layer by applying external high pressure. Large-scale negative structures of lotus leaves were obtained on polystyrene layers.

**2. Experimental details**

The natural lotus leaves were used as templates. A piece of fresh lotus leaf was fixed onto a silicon wafer by a double-sided adhesive tape. Inorganic thin films (e.g., copper and silicon) were deposited onto the surface of lotus leaves using commercial ion beam deposition system (FDJ600, Shenyang Vacuum Equipment Factory, China). A 3 cm diameter Kaufman ion source under an incidence of $45^o$ to the target provided Ar ion for sputtering. The substrate was located parallel to the target holder. The pressure during deposition was $2.4 \times 10^{-2}$ Pa. The gas flow rate was 8.8 sccm for the Ar ion beam. The Ar ion beam current was held to 50 mA. Except heating from ion beam bombardment during deposition, no external heating of the substrate was used. The deposition temperature was below 50 $^o$C as measured by a thermocouple from the backside of the substrate. The thickness of deposited inorganic films was controlled by choosing different deposition time. Then, fresh lotus leaves and inorganic films deposited on lotus leaves were used as stamps for nanoimprint lithography.

Polystyrene (PS) with a low glass transition temperature was used as nanoimprint resisting layer. PS was prepared following the reported procedure [12, 13]. PS films were spin-coated with a rotating speed of 1000 rpm from an aqueous solution onto silicon wafers cleaned by standard steps. Five layers of PS films were spin-coated on the silicon wafer and then baked in a vacuum oven at 90 °C for 30 min. The thickness of PS film was approximately 600 nm as measured by atomic force microscope (SPA400, NSK STD., Japan).

The nanoimprint processes were carried out using a commercial Nanoimprinter (NIL 2.5 in., Obducat AB, Malmoe, Sweden). Fig. 1 shows the main stages of deposition and nanoimprint procedures. The surface topography of fresh lotus leaf and lotus-leaf-like copper and silicon films before and after imprint process was investigated by scanning electron microscopy (JSM-5600LV, JEOL LTD., Japan) at 20 kV. The static water contact angle was measured with an optical contact angle meter (Dropmaster 300, Kyowa Interface Science, Japan) at room temperature. The volume of the individual water droplet used for the static contact angle measurements was 4 $\mu$L. Contact angle values were obtained by averaging five measurements results on different surface areas of the sample.

### 3. Results and discussion

Lotus leaf is famous for its superhydrophobicity and self-cleaning properties because it has microsized papillae structure and nanosized epicuticula wax on the surface. Fig. 2a and Fig. 2b present the SEM images of typical surface morphology of fresh lotus leaves. As shown in the figures, the surfaces of the natural leaves covered with micrometer-scale pillars of 3-11 $\mu$m diameters and 7-13 $\mu$m heights and branch-like wax nanostructures of about 100 nm in diameter. The microsized pillars ensured that the surface contact area available to water was small enough while the nanostructures enhanced the surface roughness and provided a water-repellent layer. We measured the wettability of natural lotus leaves and obtained the static water contact angle of about 160 ° on the surface of fresh lotus leaves, which is due to the combination of the pillars and the wax on the surface [14].

Nanoimprint process was performed to obtain large-scale replication of the surface structures of lotus leaves. Fresh lotus leaves were used directly as nanoimprint stamps. Fig. 2c and Fig. 2d show the surface morphology of fresh lotus leaf peeled off the surface of polystyrene resist layer after the nanoimprint process. The micron-sized papilla structures of lotus leaves were collapsed because of their low pressure resistance.

The nanoimprint experiments using fresh lotus leavers as stamps indicated that fresh lotus leaf could not be used directly as stamp for its natural frangibility. To be used as nanoimprint stamps, the surface structure of natural lotus leaf should be reinforced.

Inorganic thin films of copper or silicon were deposited onto fresh lotus leaves using ion beam sputtering with controlling depositional conditions. Fig. 3 shows the typical SEM images of copper and silicon films on lotus leaves templates. As shown in the figures, the inorganic films kept the micro-scale papillary structure originated from lotus leaves templates. The nanoscale features of wax had been covered by the deposited inorganic particles of about 100-200 nm in size. The density of micro-pillars on the copper deposited film (Fig. 3a and b) is lower than that of silicon film (Fig. 3c and d). The larger molecular weight of copper caused a thicker deposited copper film on lotus leaf is thicker than the silicon film within the same sputtering time and a part of smaller micrometer-pillars on the lotus leaf had been already covered.

Hydrophobic properties of the deposited inorganic films were characterized by static water contact angle (CA) measurements. The measured CA for the pure copper film on the flat silicon wafer ($92.5 \pm 4.4°$, Fig. 4a) showed a hydrophobic behavior. For the copper film coated on lotus leaves, water CA result ($136.3 \pm 8°$, Fig. 4b) exhibited the enhanced hydrophobic behavior due to its lotus-leaf-like microsized pillars structure. The CA value of pure silicon film on flat silicon substrate ($57 \pm 0.8°$, Fig. 4c) indicated a relative hydrophilic behavior. However, the CA value ($117.8 \pm 4.4°$, Fig. 4d) of the silicon film deposited on lotus leaf exhibited a crossover from a hydrophilic to a hydrophobic behavior due to the lotus-leaf-like structure. The contact angle value of silicon film increased more than copper film as a result of having large quantity of micro-pillars on silicon surface as mentioned above.

Nanoimprint experiments were carried out using the lotus-leaf-like inorganic hard stamps and PS resist layer. The optimized imprint process was performed as follows: imprint pressure was kept $30 \times 10^5$ Pa, and imprint temperature was kept 150 $^o$C for 200 s, which is higher than the glass transition temperature of PS layer, then the temperature was allowed to cool to 90 $^o$C and the hard stamp was peeled off the PS layer. Figure 5 shows the SEM images of the surface morphology of imprinted PS resist layer after thermal nanoimprint lithography using the stamps of copper (Fig. 5a) and silicon films (Fig. 5c ) on lotus leaves. As shown in the figures, imprinted microcavities appeared on the smooth PS layer with about 3-10 μm in diameter. The microcavities on the imprinted PS layer were the negative structure of micro-sized pillars on the inorganic film stamps.

The PS layers imprinted by silicon films stamps had more and deeper microcavities than those imprinted by copper films. The reasons can be explained according to the SEM observations showed in Fig. 3. Copper film stamps on the lotus leaves had an average lower height and lower density of micro-pillars compared to the silicon films stamps. Water contact angle measurements of the PS resist layer before and after imprint process were carried out. The smooth PS resist layer before imprint process exhibited a mean contact angle of 91°. The water contact angle values of flat PS films have been reported to be between 86° and 93° according to the preparation process [15-16]. A water contact angles of 98° was obtained on the imprinted PS films. The increased contact angle was due to the increased surface roughness of the presence of microcavities on the PS layer.

The closer SEM observations (Fig. 5b, 5d) also revealed that the surface of PS film was smooth without larger sticking of the polymer resist and obvious remove of the PS layer from the silicon substrate after the release of the hard stamp. The imprinted PS resist layers can be used to obtain positive replica of lotus-leaf-like structures by nanocasting or other deposition methods. The microcavities on the PS layer were also supposed to be used as containers or reactors to perform the chemical reactions in the restricted volume.

Figure 6 presents SEM images of surface morphology of hard stamps of copper film (Fig. 6a, 6b) and silicon film (Fig. 6c, 6d) peeled off the PS resist layer after thermal nanoimprint process. Compared to surface structures of as-prepared inorganic film stamps as showed in Figure 3, the height and diameter of micro-pillars had no change after nanoimprint procedure. We did not find the obvious sticking of PS layer on the stamps. The nanoimprint results showed that the stamp of inorganic coatings on lotus leaves could be used repeatedly without structural damage. The microscopic observations of the imprinted results of PS resist layer and hard stamps indicated an easy and clean separation behavior after the nanoimprint process.

The stamps used for imprint lithography typically have a high density of protrusion features and a close contact with the imprint polymer. After the imprinting process the stamp has to be removed from the substrate. A bad separation result of stamp and polymer resist layer can lead to strong adhesion of the imprinted polymer to the mold and damage of patterned structures on resist layer. This effect can easily be seen by the large areas of remove of the resist layer and obvious sticking of the polymer to the stamp after separation process. The stamps usually need careful hydrophobic treatments with deposition of the anti-stick layer, such as a thin fluorinated self-assembled monolayer on silicon based stamps, to ensure an easy and defect-free separation of the stamp and substrate during nanoimprint lithography [17]. In the nanoimprint experiments, no removal of the PS layer from the silicon substrate and adhered polymer on the stamp surface were found in the SEM observations. The hydrophobic properties (CA of about 140° for copper film) of the inorganic film stamps ensured an easy and clean separation process without any additional hydrophobic treatment of anti-stick layer to the stamps.

## 4. Conclusion

Inorganic thin films (copper, silicon) were prepared by direct deposition onto natural lotus leaves using ion beam sputtering. The deposition of inorganic films led to an enhanced lotus-leaf-like structure and resulting improved hydrophobicity compared to the inorganic film on flat silicon. Nanoimprint results showed the inorganic films were suitable for repeated hard stamps in the imprint lithography process.

**Acknowledgments**

This work was supported by the National Natural Science Foundation of China (Grant No. 20903034, 10874040) and the Cultivation Fund of the Key Scientific and Technical Innovation Project, Ministry of Education of China (Grant No. 708062).

FIGURE CAPTIONS

**Figure 1.** The depositing and nanoimprint lithography procedures. (a): pure lotus leaf. (b): depositing inorganic film on lotus leaf using ion beam sputtering. (c): spin-coating PS film onto silicon wafer cleaned by standard steps. (d): nanoimprint procedure. (e): separation of template and resist layer.

**Figure 2.** SEM images of the natural lotus leaves (a-b) before and (c-d) after the imprint process. Angle of view: 60°.

**Figure 3.** SEM images of surface topography after depositing (a, b) copper and (c, d) silicon on the lotus leaf. Angle of view: 60°.

**Figure 4.** Water contact angle measurements on (a) copper film deposited on silicon wafer, (b) as-prepared lotus-leaf-like copper film, (c) silicon film deposited on silicon wafer (d) as-prepared lotus-leaf-like silicon film.

**Figure 5.** SEM images of the PS film imprinted with (a, b) copper films and (c, d) copper films coated on lotus leaves after nanoimprint process.

**Figure 6.** SEM images of the (a, b) copper films and (c, d) silicon films stamps after nanoimprint process. Angle of view: 60°.

**Fig.1**

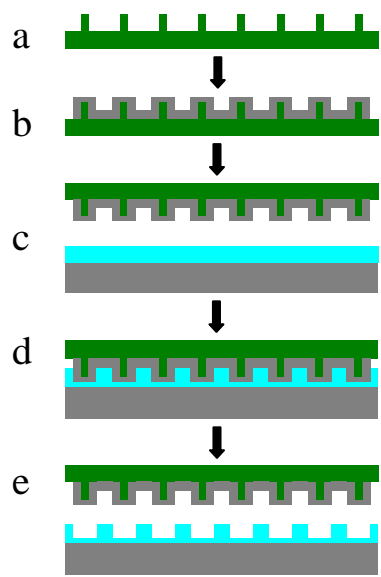

**Fig. 1** The depositing and nanoimprint lithography procedures. (a): pure lotus leaf. (b): depositing inorganic film on lotus leaf using ion beam sputtering. (c): spin-coating PS film onto silicon wafer cleaned by standard steps. (d): nanoimprint procedure. (e): separation of template and resist layer.

**Fig.2**

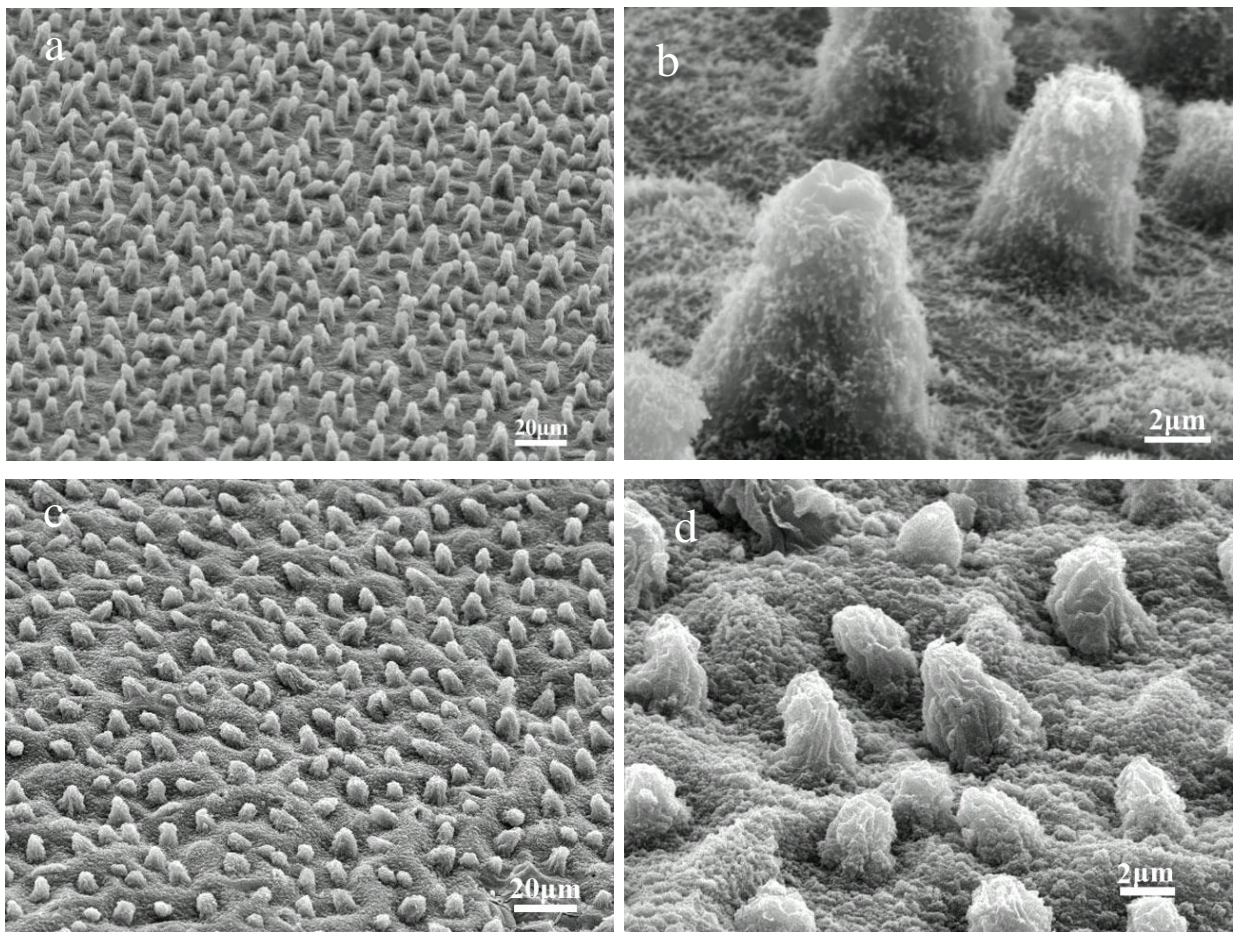

**Fig. 2** SEM images of the natural lotus leaves (a-b) before and (c-d) after the imprint process. Angle of view: 60 °.

**Fig.3**

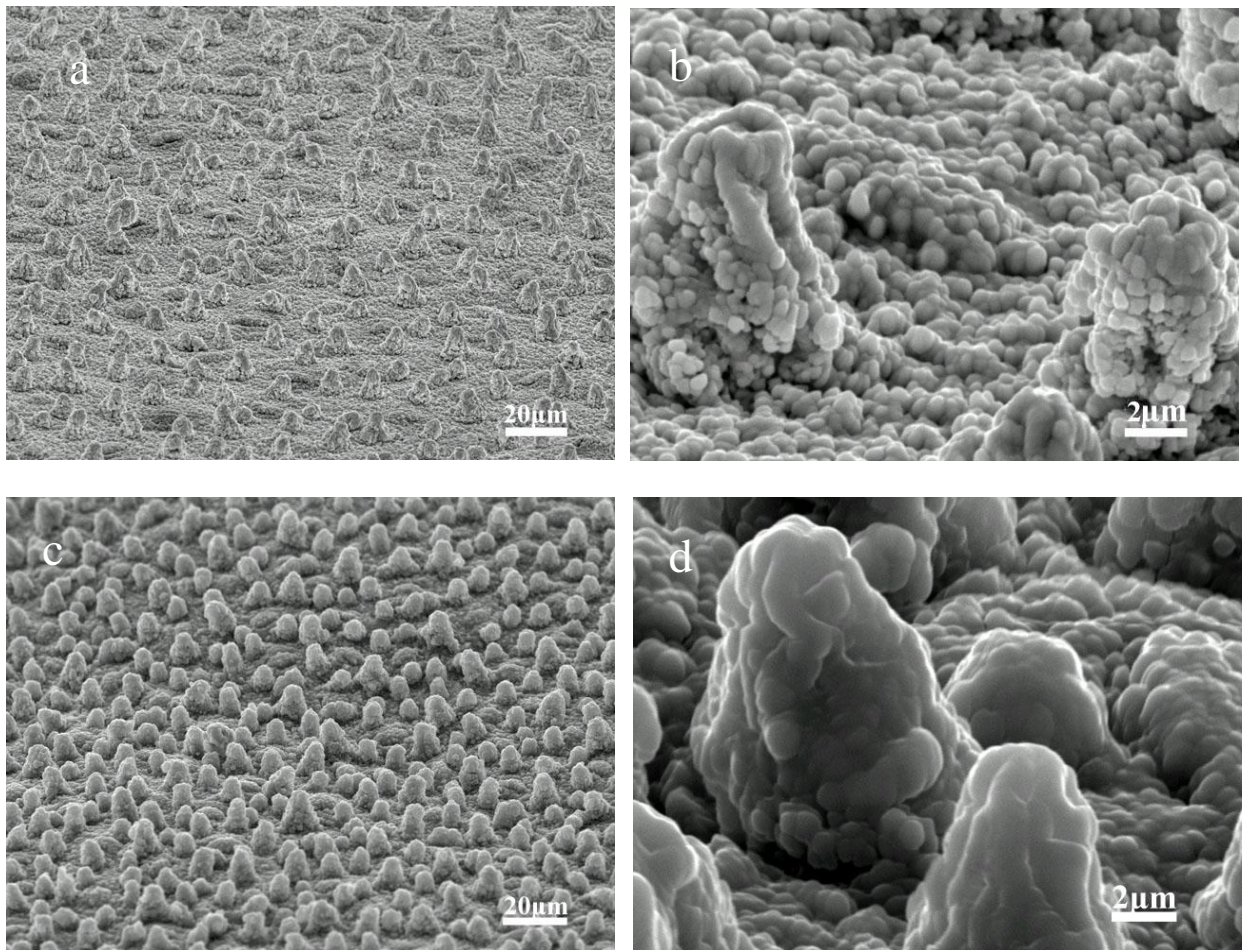

**Fig. 3** SEM images of surface topography after depositing (a, b) copper and (c, d) silicon on the lotus leaves. Angle of view: 60 °.

**Fig.4**

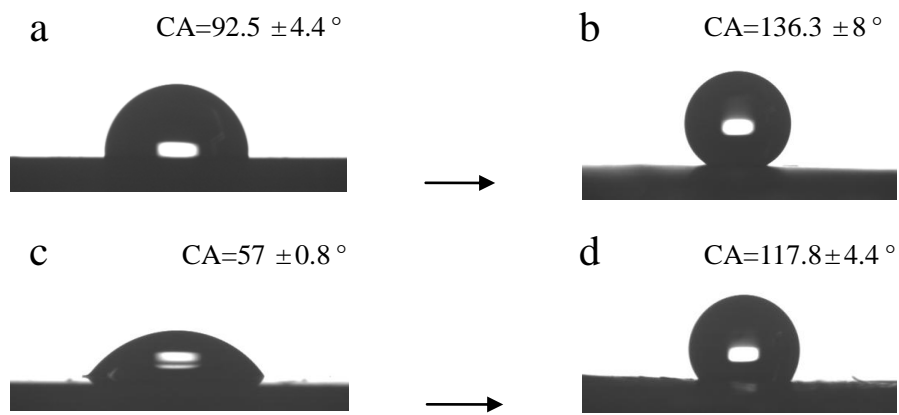

a  CA=92.5 ±4.4°
b  CA=136.3 ±8°
c  CA=57 ±0.8°
d  CA=117.8±4.4°

**Fig. 4** Water contact angle measurements on (a) copper film deposited on silicon wafer, (b) as-prepared lotus-leaf-like copper film, (c) silicon film deposited on silicon wafer (d) as-prepared lotus-leaf-like silicon film.

**Fig.5**

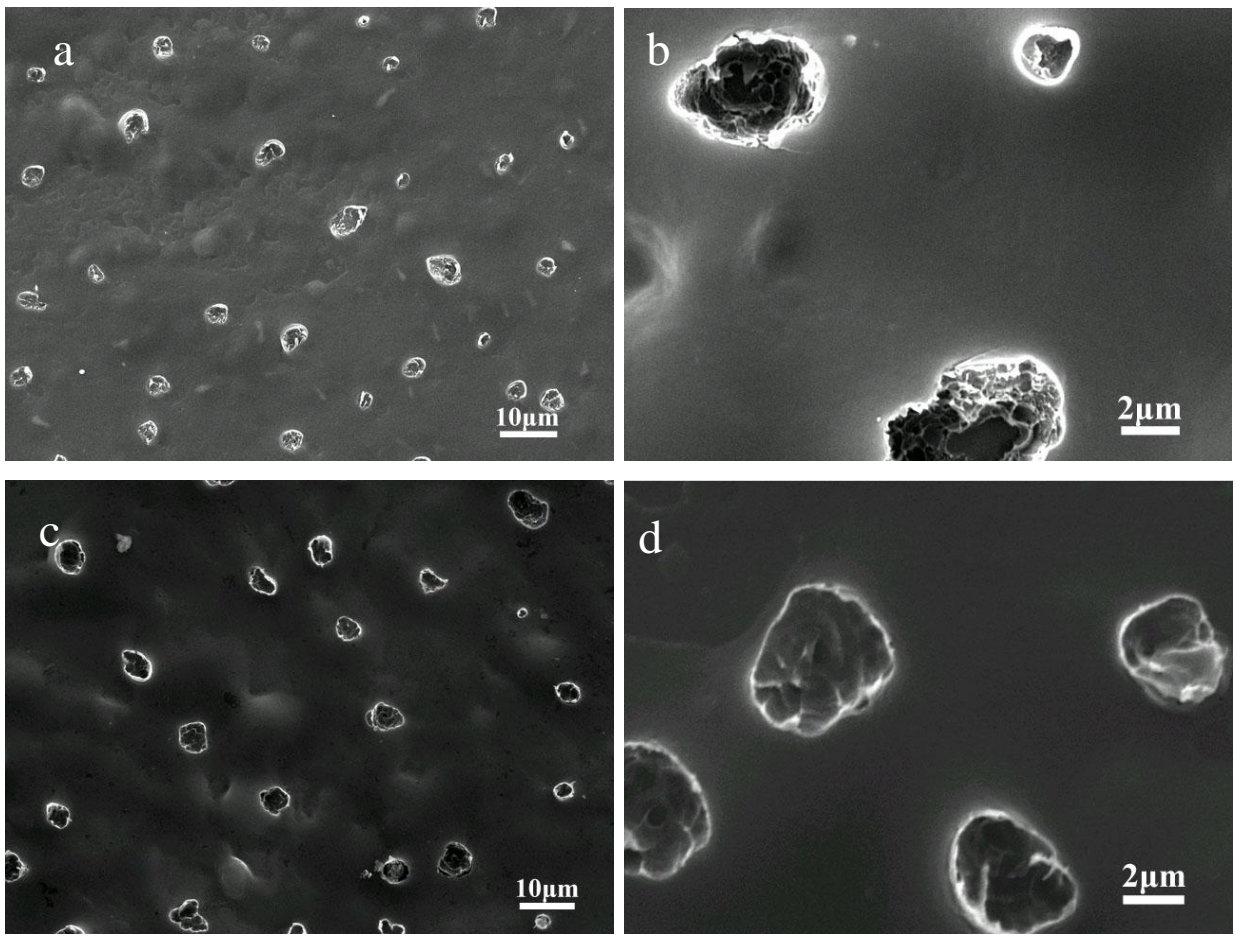

**Fig. 5** SEM images of the imprinted PS resist layer with (a, b) copper films and (c, d) silicon films coated on lotus leaves after nanoimprint process.

**Fig.6**

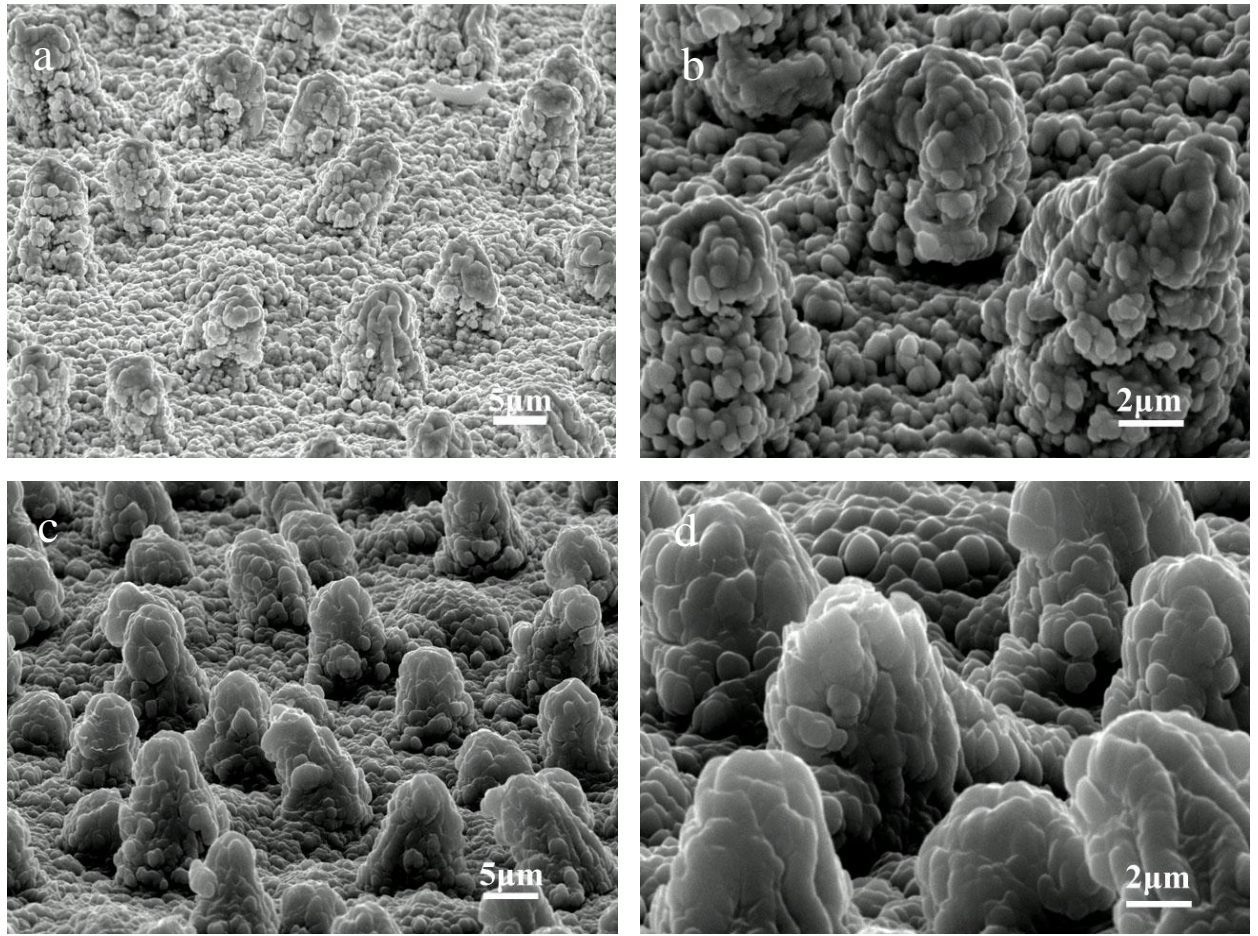

**Fig. 6** SEM images of the (a, b) copper films and (c, d) silicon films stamps after nanoimprint process. Angle of view: 60 °.